\documentclass[12pt]{elsart}

\usepackage{amsmath}
\usepackage{amsfonts}
\usepackage{graphicx}

\begin{document}

\begin{frontmatter}

\title{Alternative evaluation of \\
statistical indicators in atoms: \\
the non-relativistic and relativistic cases}

\author[jsr]{Jaime Sa\~{n}udo} and
\ead{jsr@unex.es}
\author[rlr]{Ricardo L\'{o}pez-Ruiz}
\ead{rilopez@unizar.es}

\address[jsr]{
Departamento de F\'isica, Facultad de Ciencias, \\
Universidad de Extremadura, E-06071 Badajoz, Spain, \\
and BIFI, Universidad de Zaragoza, E-50009 Zaragoza, Spain}

\address[rlr]{
DIIS and BIFI, Facultad de Ciencias, \\
Universidad de Zaragoza, E-50009 Zaragoza, Spain}


\begin{abstract}
In this work, the calculation of a statistical measure of complexity and the Fisher-Shannon 
information is performed for all the atoms in the periodic table. 
Non-relativistic and relativistic cases are considered. 
We follow the method suggested in 
[C.P. Panos, N.S. Nikolaidis, K. Ch. Chatzisavvas, C.C. Tsouros, arXiv:0812.3963v1]
that uses the fractional occupation probabilities of electrons in atomic orbitals,
instead of the continuous electronic wave functions. 
For the order of shell filling in the relativistic case, 
we take into account the effect due to electronic spin-orbit interaction.
The increasing of both magnitudes, the statistical
complexity and the Fisher-Shannon information, with the atomic number $Z$ is observed.
The shell structure and the irregular shell filling is well displayed by the Fisher-Shannon
information in the relativistic case.
\end{abstract}


\end{frontmatter}

\maketitle

In the last years, the application of different information-theoretic magnitudes, 
such as Shannon, Fisher, R\'enyi, and Tsallis entropies and statistical complexities, 
in quantum systems has taken a growing interest \cite{gadre2003,panos2005}. 
Particularly, the use of these indicators in the study of the electronic structure of atoms
has received special attention \cite{gadre1985,panos2007,sen2007,borgoo2007,angulo2008,romera2008}.

The basic ingredient to calculate these statistical magnitudes is the electron probability density,
$\rho(\vec{r})$, that can be obtained from the numerically derived Hartree-Fock atomic wave function
in the non-relativistic case \cite{panos2005,panos2007}, 
and from the Dirac-Fock atomic wave function in the relativistic case \cite{borgoo2007}.
The behavior of these statistical quantifiers with the atomic number $Z$ has revealed a
connection with physical measures, such as the ionization potential and the static dipole 
polarizability \cite{sen2007}. All of them, theoretical and physical magnitudes,
are capable of unveiling the shell structure of atoms, specifically the closure of shells 
in the noble gases. Also, 
it has been observed that statistical complexity fluctuates around an average 
value that is non-decreasing as the atomic number $Z$ increases in the non-relativistic 
case \cite{panos2007,borgoo2007}.
This average value becomes increasing in the relativistic case \cite{borgoo2007}.
This trend has also been confirmed when the atomic electron density is obtained
with a different approach \cite{sanudo2008}.
 
An alternative method to calculate the statistical magnitudes can be used when the atom
is seen as a discrete hierarchical organization. The atomic shell structure can also be captured
by the fractional occupation probabilities of electrons in the different atomic orbitals.
This set of probabilities has been employed in \cite{panos2009} to evaluate all these quantifiers
for the non-relativistic ($NR$) case.
On one hand, a non-decreasing trend in complexity as $Z$ increases is recovered.
On the other hand, the closure of shells for some noble gases is now masked. 

In order to complement the results given in \cite{panos2009} for the $NR$ case, 
here we undertake the calculation for the relativistic ($R$) case by also using the fractional 
occupation probabilities of electrons in atomic orbitals.
 
For the $NR$ case, each electron shell of the atom is given by $(nl)^w$ \cite{bransden2003}, 
where $n$ denotes the principal quantum number, $l$ the orbital angular momentum $(0\leq l\leq n-1)$ 
and $w$ is the number of electrons in the shell $(0\leq w\leq 2(2l+1))$. 
For the $R$ case, due to the spin-orbit interaction, 
each shell is split, in general, in two shells: 
 $(nlj_-)^{w_-}$, $(nlj_+)^{w_+}$, where $j_{\pm}=l\pm 1/2$ 
 (for $l = 0$ only one value of $j$ is possible, $j=j_+=1/2$) and
 $0\leq w_{\pm}\leq 2j_{\pm}+1$ \cite{cowan1981}. 
 As an example, we explicitly give the electron configuration of  
 $Ar(Z=18)$ in both cases,
 \begin{eqnarray}
 Ar(NR) & : & (1s)^2(2s)^2(2p)^6(3s)^2(3p)^6, \\
 Ar(R)\;\;\; & : & (1s1/2)^2(2s1/2)^2(2p1/2)^2(2p3/2)^4(3s1/2)^2(3p1/2)^2(3p3/2)^4. 
 \end{eqnarray}

For each atom, a fractional occupation probability distribution of electrons 
in atomic orbitals $\{p_k\}$, $k = 1,2,\ldots,\nu$,  being $\nu$ the number of shells of the atom,
can be defined. This normalized probability distribution $\{p_k\}$ $(\sum p_k=1)$ is easily calculated by dividing 
the superscripts $w_{\pm}$ (number of electrons in each shell) by $Z$, the total number of electrons in neutral atoms, 
which is the case we are considering here. It should also be mentioned that the order of shell filling
dictated by nature \cite{bransden2003} has been chosen. Then, from this probability distribution,
the different statistical magnitudes (Shannon entropy, disequilibrium, statistical complexity and
Fisher-Shannon entropy) can be calculated.

In this work, we calculate the so-called LMC complexity \cite{lopez1995,lopez2002}, 
a statistical measure of complexity, $C$, that has been recently used 
to enlighten different questions on the hierarchical organization
of some few-body quantum systems \cite{sanudo2008-,sanudo2008+,sen2008,howard2008} and 
also in the case of many-body quantum systems 
\cite{panos2005,panos2007,borgoo2007,angulo2008,panos2009}. 
It is defined as 
\begin{equation}
C_{LMC} = H\cdot D\;,
\end{equation}
where $H$, that represents the information content of the system, 
is in this case the simple exponential Shannon entropy \cite{lopez2002,dembo1991},
\begin{equation}
H = e^{S}\;,
\end{equation}
$S$ being the Shannon information entropy \cite{shannon1948},
\begin{equation}
S = -\int \rho(x)\;\log \rho(x)\; dx \;,
\label{eq1}
\end{equation}
where $\rho$  is the electron density normalized to unity. $D$ gives an idea of how much 
concentrated is its spatial distribution and it is calculated as the density expectation 
value \cite{lopez1995,lopez2002}
\begin{equation}
D = \int \rho^2(x)\; dx\;.
\label{eq2} 
\end{equation}
The discrete versions of expressions (\ref{eq1}) and (\ref{eq2}) used in our calculations
are given by 
\begin{eqnarray}
\hspace{4.3cm} S & = & -\sum_{k=1}^{\nu}p_k\log p_k \; , \\
\hspace{4.3cm} D & = & \;\;\sum_{k=1}^{\nu}(p_k-1/\nu)^2 \; .
\end{eqnarray}

The Fisher-Shannon information, $P$, has also been applied in atomic 
systems \cite{sanudo2008-,sanudo2008+,sen2008,howard2008,romera2004,szabo2008}. 
This quantity is given by
\begin{equation}
P = J\cdot I \; ,
\label{eq-p}
\end{equation}
where $J$ is a version of the power Shannon entropy \cite{dembo1991}
\begin{equation}
H = {1\over 2\pi e}\; e^{2S/3}\;,
\end{equation}
whereas $I$ is the so-called Fisher information measure \cite{fisher1925}, 
that quantifies the narrowness of the probability density and it is given by
\begin{equation}
I = \int {|\nabla\rho(\vec{r})|^2\over \rho(\vec{r})}\; d\vec{r}\;.
\label{eq-I}
\end{equation}
In order to apply the present method, the same discrete version of $I$ 
as in \cite{panos2009} is used 
\begin{equation}
I = \sum_{k=1}^{\nu} {(p_{k+1}-p_k)^2\over p_k}\;,
\end{equation}
where $p_{\nu+1}=0$.

\begin{figure}[h]  
\centerline{\includegraphics[width=12cm]{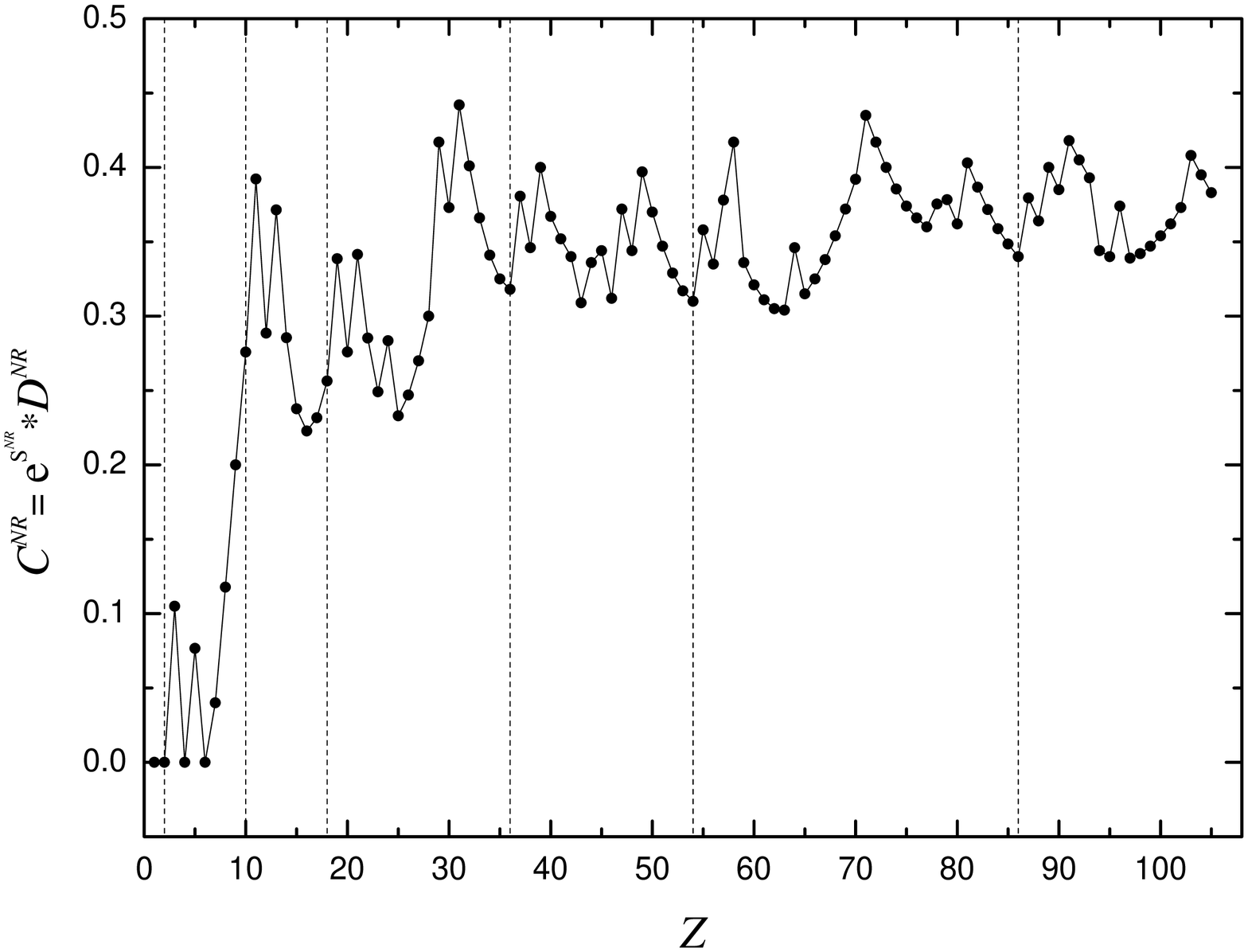}}  
\caption{Statistical complexity, $C$, vs. $Z$ in the non relativistic case ($C^{NR}$). 
The dashed lines indicate the position of noble gases. For details, see the text.}  
\label{fig1}  
\end{figure}

\begin{figure}[h]  
\centerline{\includegraphics[width=12cm]{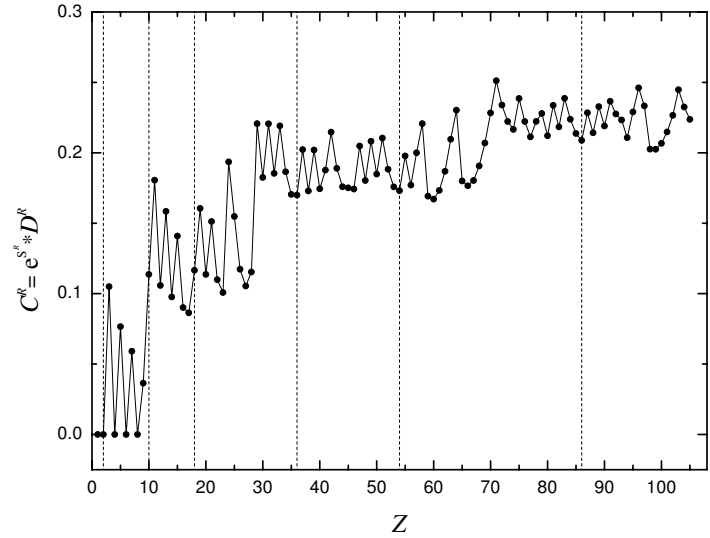}}     
\caption{Statistical complexity, $C$, vs. $Z$ in the relativistic case ($C^R$). 
The comments given in Fig. \ref{fig1} are also valid here.}  
\label{fig2}  
\end{figure}

The statistical complexity, $C$, as a function of the atomic number, $Z$, 
for the $NR$ and $R$ cases for neutral atoms is given in Figs. \ref{fig1} and \ref{fig2}, 
respectively. 
We can observe in both figures that this magnitude fluctuates around an increasing average value 
with $Z$. This increasing trend recovers the behavior obtained in \cite{panos2007,borgoo2007}
by using the continuous quantum-mechanical wave functions, although it is not the same 
as found in \cite{panos2009}. This different tendency can be explained by the fact that
we have used $H=e^S$, whereas in \cite{panos2009} the authors take $H=S$.
A shell-like structure is also unveiled in this approach by looking at the minimum values of $C$
taken on the noble gases positions (the dashed lines in the figures) with the 
exception of $Ne(Z=10)$ and $Ar(Z=18)$. 

\begin{figure}[h]  
\centerline{\includegraphics[width=12cm]{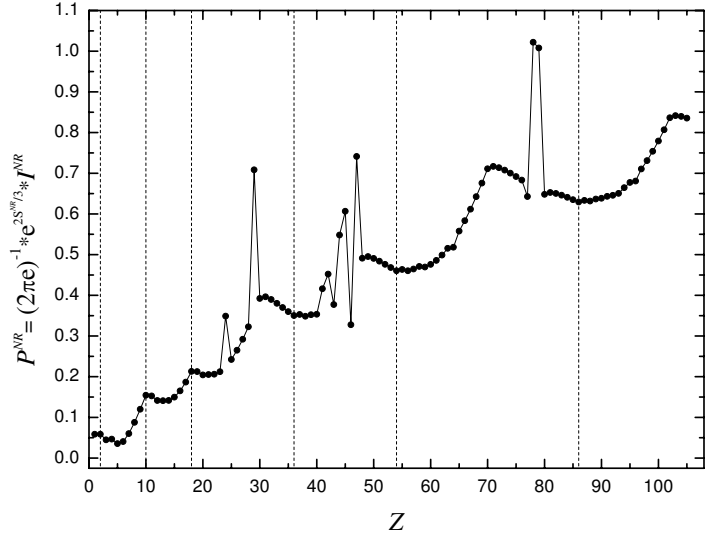}}     
\caption{Fisher-Shannon entropy, $P$, vs. $Z$, in the non relativistic case ($P^{NR}$). 
The dashed lines indicate the position of noble gases. For details, see the text.}  
\label{fig3}  
\end{figure}

\begin{figure}[h]  
\centerline{\includegraphics[width=12cm]{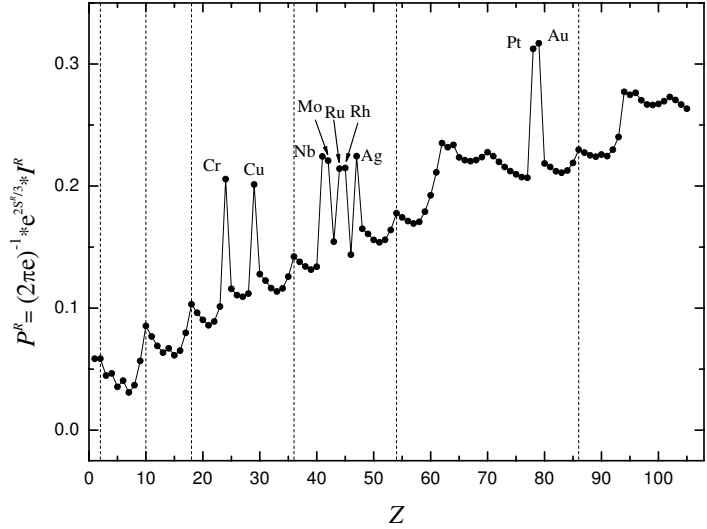}}     
\caption{Fisher-Shannon entropy, $P$, vs. $Z$, in the relativistic case ($P^R$). 
The comments given in Fig. \ref{fig3} are also valid here.}  
\label{fig4}  
\end{figure}

The Fisher-Shannon entropy, $P$, as a function of $Z$, for the $NR$ and $R$ cases in neutral 
atoms is given in Figs. \ref{fig3} and \ref{fig4}, respectively. 
The shell structure is again displayed, especially in the $R$ case (Fig. \ref{fig4})
where  $P$ takes local maxima for all the noble gases (see the dashed lines on $Z = 2, 10, 18, 36, 54, 82$).
The irregular filling (i.f.) of $s$ and $d$  shells \cite{bransden2003} is also detected by peaks
in the magnitude $P$, mainly in the $R$ case. In particular, see the elements $Cr$ and $Cu$  
(i.f. of $4s$ and $3d$ shells); $Nb$, $Mo$, $Ru$, $Rh$, and $Ag$ (i.f. of $5s$ and $4d$ shells); 
and finally $Pt$ and $Au$ (i.f. of $6s$ and $5d$ shells). 
$Pd$ also has an irregular filling, but $P$ does not display a peak on it
because the shell filling in this case does not follow the same procedure 
as the before elements (the $5s$ shell is empty and the $5d$ is full). 
Finally, the increasing trend of $P$ with $Z$ is clearly manifested.

Then, we conclude that if the fractional occupation probabilities of electrons in atomic orbitals,
instead of the continuous electronic wave functions, are used to calculate $C$ and $P$,
it is found that $P$, the Fisher-Shannon entropy, 
in the relativistic case (Fig. \ref{fig4}) reflects in a better way the increasing trend with $Z$,
the shell structure in noble gases, and the irregular shell filling of some specific elements.

\newpage  
  
\newpage  

\end{document}